\documentstyle[aps,12pt,preprint]{revtex}
\def \to {\rightarrow}
\def \beq {\begin{equation}}
\def \eeq {\end{equation}}
\def \ba {\begin{eqnarray}}
\def \ea {\end{eqnarray}}
\def \jpsi {J/\psi}
\def \< {\left <}
\def \> {\right >}

\begin{document}
\baselineskip 20pt
\renewcommand{\thesection}{\Roman{section}}
~
{\hfill BUTP-96-36}
\vskip 20mm
\begin{center}
{\Large \bf $D$-wave charmonium production in $B$ decays}
\end{center}
\vskip 10mm
\centerline{Feng Yuan$^a$,~~Cong-Feng Qiao$^b$~~and~~Kuang-Ta Chao$^{a,b}$}
\vskip 2mm
\centerline{\small\it $^b$Department of Physics, Peking University, 
            Beijing 100871, People's Republic of China}
\vskip 1mm
\centerline{\small\it $^a$China Center of Advanced Science and Technology
(World Laboratory), Beijing 100080, People's Republic of China}
\vskip 15mm

\begin{center}
{\bf\large Abstract}

\begin{minipage}{140mm}
\vskip 5mm
\par

The calculation of $D$-wave charmonium prodution rates in B meson deacys under
the NRQCD factorization formalism is presented. We find that
inclusion of the color-octet contributions permits us to detect the 
$D$-wave charmonium states in $B$ decays at present experimental facilities.
The same amount signals of $2^{--}$ $D$-wave state as that of $\psi^\prime$
could be observed at {\bf LEP} and {\bf CESR}.
We also predict the relative production rates for four $D$-wave states are
$\delta^c:\delta^c_1:\delta^c_2:\delta^c_3=2.5:3:5:7$, where $\delta^c,~
\delta^c_1,~\delta^c_2,~\delta^c_3$ represent respectively the $2^{-+},~
1^{--},~2^{--},~3^{--}$ states.

\vskip 5mm
\noindent
PACS number(s): 12.25.Hw, 12.38.Lg, 14.40.Nd
\end{minipage}
\end{center}
\vfill\eject\pagestyle{plain}\setcounter{page}{1}

\section{Introduction}

Studies of heavy quarkonium production in high energy collisions provide important
imformation on both perturbative and nonperturbative QCD.
In recent years, a rigorous framwork for treating quarkonium production 
and decays has been advocated by Bodwin, Braaten and Lepage in the context of 
nonrelativistic quantum chromodynamics (NRQCD)\cite{bbl}. 
In this approach, the production process is factorized into short and long 
distance parts, while the latter is associated with the nonperturbative
matrix elements of four-fermion operators. This factorization formalism
provides a new production mechanism called the color-octet mechanism, in which
the heavy-quark and antiquark pair is produced at short distance in a
color-octet configuration and subsequently
evolves nonperturbatively into physical quarkonium state.
This mechanism is first considered to cancel the infrared divergence
in the calculation of $P$-wave charmonium production in $B$ decays\cite{b-pwave}.
But the most important progress is that by including the color-octet production
mechanism one might explain the $\psi^\prime$ ($\jpsi$) surplus
measured by {\bf CDF} at the {\bf Tevatron}\cite{surplus}.
In the past few years, applications of the NRQCD factorization formalism
to $\jpsi$($\psi^\prime$) production at various experimental facilities
have been studied\cite{annrev}.

Recently, a calcultion of $D$-wave charmonium production in $Z^0$ decays suggests
a crucial test of color-octet production mechanism\cite{z0}. The authors show
that the production rates due to color-octet gluon fragmentation to $D$-wave
charmonium states are $2\sim 3$ orders larger than that due to the dominant
color-singlet quark fragmentation\cite{quark}.
They also predict that at the {\bf Tevatron} the same amount of magnitude
of direct prodution rate of $2^{--}$ $D$-wave charmonium state as
$\psi^\prime$ could be observed\cite{gluon}.
All these progresses show that the color-octet mechanism is crucial 
important to $D$-wave charmonium production
because the color-octet contributions are over 
two orders larger than the color-singlet contributions.
While in $S$- or $P$-wave states production, the situations are not so critical.
This huge divergences between color-octet and color-singlet contributions
in $D$-wave states production are most helpful in distinguishing the two production
mechanisms in experiments and will provide a crucial test of color-octet production mechanism. 
On the other hand, even if the color-singlet model predicts the $D$-wave production rates
too small to be visible, they could now be detected after and only after
including the color-octet production mechanism.
So it is the color-octet mechanism that could make it possible to search for the $D$-wave
heavy quarkonium states, and then will complete the studies of
charmonium and bottomonium families.

In this paper, we will calculate $D$-wave charmonium production rates in $B$ decays,
as a complementation to previous studies\cite{z0}\cite{gluon}, and also
as an independent determination of the color-octet matrix elements
associated with color-octet $D$-wave charmonium production.
Color-singlet $S$- and $P$-wave charmonium prodcution in $B$ decays 
have been studied in the literatures\cite{singlet}. The color-octet
$S$- and $P$-wave production have also been investigated\cite{b-pwave}\cite{b-swave}\cite{matrix}.
>From the following calculations, we will show that the color-octet production
mechanism is also crucial important to $D$-wave charmonium production in
$B$ decays.
The rest of the paper is organized as follows. In Set.II, we give the factorization
formulas under the NRQCD formalism.
In Sec.III, we estimate the branching fractions of $D$-wave charmonium production
in $B$ decays by assuming that the values of color-octet matrix elements
are taken by the NRQCD velocity scaling rules.
We also disccuss the detection of $D$-wave charmonium states
in B deacys at present experimental facilities in this section.
A short summary will be given in Sec.IV.

\section{Factorization formula}

According to the NRQCD factorization formalism, the patial width for the
inclusive production
of a given charmonium state $H$ in $b$ decays has the following factorization form,
\beq
\label{expansion}
\Gamma (b\to H+X)=\sum\limits_n\hat{\Gamma}(b \to c\bar c(n) +X)
        \< {\cal O}_n^H \> .
\eeq
Here, $\hat\Gamma_n$ is the short-distance subprocess rate for producing a $c\bar c$
pair in configuration denoted by $n$ (including the angular momentum $^{2S+1}
L_J$ and the color index $1$ or $8$). $\< {\cal O}_n^H \> $ is the long-distance
nonperturbative matrix element demonstrating the probability of 
the $c\bar c$ pair in $n$ configuration evolving into a physical charmonium
state $H$.
While $\hat\Gamma_n$ can be calculated perturbatively as an expansion in
coupling constant $\alpha_s(M_b)$, $\< {\cal O}_n^H \> $ is a
nonperturbative parameter, and practically can only be determined by fitting to the
experimental data. Fortunately, under the framework of NRQCD, the relative size of
$\< {\cal O}_n^H \> $ can be roughly estimated by using its scaling property in $v^2$ (controled by
the velocity scaling rules), where $v$ is the typical relative velocity of
the heavy quark inside the bound state $H$.

The short-distance subprocesses $b\to c\bar c q$ (with $q=s,d$) are described
by the effective Hamitonian\cite{b-pwave}
\ba
\label{hamiton}
\nonumber
H_{eff}&=&-\frac{G_F}{\sqrt{2}}V_{cb}V_{cs}^*\big ( \frac{2C_+-C_-}{3}
        \bar c\gamma_\mu (1-\gamma_5)c\bar s \gamma_\mu (1-\gamma_5)b\\
        &~&+
        (C_++C_-)\bar c\gamma_\mu(1-\gamma_5)T^a c \bar s\gamma^\mu(1-\gamma_5)
        T^a b\big ),
\ea
where $G_F$ is the Fermi constant and $V_{ij}$'s are KM matrix elements.
The coefficiets $C_+$ and $C_-$ are Wilson coefficients at the scale of $\mu=M_b$.
To leading order of $\alpha_s(M_b)$ and to all orders of $\alpha_s(M_b)ln(M_W/M_b)$,
they are
\ba
C_+(M_b)&\approx& [\alpha_s(M_b)/\alpha_s(M_W)]^{-6/23},\\
C_-(M_b)&\approx& [\alpha_s(M_b)/\alpha_s(M_W)]^{12/23}.
\ea
In this effective Hamitonian Eq.(\ref{hamiton}), the first term provides the color-singlet contribution
of charmonium production in $b$ decays, {\it i.e.}, it produces $c\bar c$ pair
in a color-singlet state, while the second term produces $c\bar c$ pair in a color-octet state
and provides the color-octet contribution.

To $D$-wave charmonium production, as argued in\cite{z0}\cite{gluon}, the color-singlet
and color-cotet processes are both scaled with the same orders in $v^2$.
In $b$ decays, the color-octet contributions include ${}^1S_0^{(8)}$ subprocess
for spin-singlet and ${}^3S_1^{(8)}$ and ${}^3P_1^{(8)}$ subprocesses for
spin-triplet $D$-wave states.
In the factorization formula Eq.(\ref{expansion}), the associated matrix elements
are all in $v^7$, and they have following relations
according to the NRQCD velocity scaling rules,
\beq
\< {\cal O}_1^{\delta^c}({}^1D_2)\> \sim M_c^7 v^7,~~
\< {\cal O}_8^{\delta^c}({}^1S_0)\> \sim M_c^3 v^7;
\eeq
\ba
\< {\cal O}_1^{\delta^c_J}({}^3D_J)\> \sim M_c^7 v^7,~~
\< {\cal O}_8^{\delta^c_J}({}^3S_1)\> \sim M_c^3 v^7,~~
\< {\cal O}_8^{\delta^c_J}({}^3P_1)\> \sim M_c^5 v^7.
\ea
Here, the symbol $\delta^c$ represents the physical spin-singlet $D$-wave
charmonium state, and $\delta^c_J$ ($J=1,~2,~3$) for spin-triplet $1^{--}$,
$2^{--}$, $3^{--}$ states. The notations
${}^{2S+1}D_J$ represent the $c\bar c$ pairs configurations with angular
momentum $L=2$. Because the color-singlet and color-octet processes
are all in the same orders in $v^2$, at leading order in $\alpha_s$ and $v^2$,
all these processes must be taken into account for a consistent calculation.
The color-singlet matrix elements $\< {\cal O}_1^{\delta^c_J}({}^3D_J)\> $
can be related to the second derivative of the nonrelativistic radial wave
function at the origin $|R_D^{\prime\prime}(0)|^2$ for $D$-wave by
\ba
\< {\cal O}_1^{\delta}({}^1D_2)\> &=&\frac{75N_c}{4\pi}|R_D^{\prime\prime}(0)|^2,\\
\< {\cal O}_1^{\delta_J}({}^3D_J)\> &=&\frac{15(2J+1)N_c}{4\pi}|R_D^{\prime\prime}(0)|^2.
\ea
The wave function may be evaluated by potential model so that the color-singlet
matrix elements are fixed. Whereas the color-octet matrix elements
$\< {\cal O}_8^{\delta^c_J}({}^3S_1)\> $ and $\< {\cal O}_8^{\delta^c_J}({}^3P_1)\> $
are free parameters and must be determined by experiment.

For the spin-singlet $D$-wave charmonium state, at leading order in $\alpha_s$, the color-singlet
subprocess can not contribute to its production in $b$ decays.
At leading order in $v^2$, the production can only come from the color-octet
${}^1S_0^{(8)}$ subprocess.
Using the effective Hamitonian(\ref{hamiton}), one can calculate the short-distance
coefficient\cite{b-swave}, and get
\beq
\label{ds0}
\Gamma(b\to c\bar c({}^1S_0^{(8)})+s,d\to\delta^c+X)=
        \frac{3\< {\cal O}_8^{\delta^c}({}^1S_0)\> }{2M_c^2}(C_++C_-)^2\hat\Gamma^0,
\eeq
with
\beq
\hat\Gamma^0=|V_{cb}|^2\frac{G_F^2}{144\pi}M_b^3M_c(1-\frac{4M_c^2}{M_b^2})^2.
\eeq
Here the relation $|V_{cs}|^2+|V_{cd}|^2\approx 1$ has been used.

For spin-triplet $D$-wave states, at leading order in $\alpha_s$, the
color-singlet process can only contribute to the production of state ${}^3D_1$.
Other two states (${}^3D_2,~{}^3D_3$) can not be produced
in color-singlet process and can only  be produced in color-octet processes.

We first calculate the color-singlet contribution to ${}^3D_1$ charmonium
production in $b$ decays.
This color-singlet production rate can be calculated by making use of the 
covariant
formalism\cite{singlet}.
>From standard BS wave function in nonrelativistic approximation for vector
mesons, the bound state wave function can be project out as,
\begin{equation}
 \Phi(P,\vec{q})=\sum_{sm}\langle JM|1sLm\rangle 
         \!\not{\epsilon}(1+\frac{\!\not{P}}{M})\psi_{Lm}
         (\vec{q}).
\end{equation}
To $D$-wave function, $\Phi(P,\vec{k})$ must be expanded to the secdond order in
the relative momentum $\vec{q}$.
The first and second derivative of the wave function are
\begin{equation}
 \Phi_\alpha(\vec{q})=\frac{-1}{2 M_c^2 M}\sum_{sm}\langle JM|1sLm\rangle  
 [M_c\gamma_\alpha \!\not{\epsilon}(M+\not\!{P}) + 
 M_c\!\not{\epsilon}(M+\not\!{P})\gamma_\alpha ]
 \psi_{Lm}(\vec{q});
\end{equation}
\begin{equation}
 \Phi_{\alpha\beta}(\vec{q})=\frac{-1}{2 M_c^2 M}\sum_{sm}\langle JM|1sLm\rangle  
 \gamma_\alpha\!\not{\epsilon}(M+\not\!{P})\gamma_\beta\psi_{Lm}(\vec{q}).
\end{equation}
After integrating over the relative momentum $\vec{q}$, the orbit angular momentum
part of the wave function will depend on the radial wave function or its
derivatives at the orgin
$R_S(0)$, $R_P^\prime(0)$, and $R_D^{\prime\prime}(0)$, respectively, for
$S$-wave ($L=0$), $P$-wave ($L=1$), and $D$-wave ($L=2$) states,
\begin{equation}
 \int\frac{d^3 q}{(2\pi)^3}\psi_{00}({\bf q})=\frac{1}{\sqrt{4\pi}}R_S(0),
\end{equation}
\begin{equation}
 \int\frac{d^3 q}{(2\pi)^3}q^{\alpha}
 \psi_{1m}({\bf q})=i\epsilon^{\alpha}\sqrt{\frac{3}{4\pi}}R_P^{\prime}(0),
\end{equation}
\begin{equation}
 \int\frac{d^3 q}{(2\pi)^3}q^{\alpha}q^{\beta}
 \psi_{2m}({\bf q})=e^{\alpha\beta}_m\sqrt{\frac{15}{8\pi}}
 R_D^{\prime\prime}(0).
\end{equation}
where the polarization tensor's label $m$ is magnetic quantum number. For 
the spin-triplet case where $J=1,~2,~3$, using explicit Clebsch-Gordan coefficients, 
there are following relations for three cases \cite{Be}.
\begin{eqnarray}
\nonumber  
  \sum_{sm}\langle 1J_z|1s2m\rangle\epsilon_{\alpha\beta}^{(m)}\epsilon_\rho^{(s)}
 &=&-[{3\over 20}]^{1/2}[(g_{\alpha\rho}-\frac{p_\alpha p_\rho}{4M_c^2})\epsilon_\beta^{(J_z)}
     +(g_{\beta\rho}-\frac{p_\beta p_\rho}{4M_c^2})\epsilon_\alpha^{(J_z)}\\
  &~&-{2\over 3}(g_{\alpha\beta}-\frac{p_\alpha p_\beta}{4M_c^2})\epsilon_\rho^{(J_z)}],\\
  \sum_{sm}\langle 2J_z|1s2m\rangle\epsilon_{\alpha\beta}^{(m)}\epsilon_\rho^{(s)}
         &=&\frac{i}{2\sqrt{6}M_c}(\epsilon_{\alpha\sigma}^{(J_z)}\epsilon_{\tau\beta\rho\sigma^{\prime}}p^\tau g^{\sigma\sigma^\prime}+
             \epsilon_{\beta\sigma}^{(J_z)}\epsilon_{\tau\alpha\rho\sigma^{\prime}}p^\tau g^{\sigma\sigma^\prime}),\\
  \sum_{sm}\langle 3J_z|1s2m\rangle\epsilon_{\alpha\beta}^{(m)}\epsilon_\rho^{(s)}
         &=&\epsilon_{\alpha\beta\rho}^{(J_z)}.
\end{eqnarray}
   Here, $\epsilon_\alpha$, $\epsilon_{\alpha\beta}$, $\epsilon_{\alpha\beta\rho}$
are spin-one, spin-two and spin-three polarization tensors which obey 
the projection relations\cite{Be}
\begin{eqnarray}
\sum_m \epsilon_\alpha^{(m)}\epsilon_\beta^{(m)}&=&(-g_{\alpha\beta}+\frac{p_\alpha p_\beta}{4M_c^2})
   \equiv{\cal P}_{\alpha\beta},\\
\sum_m \epsilon_{\alpha\beta}^{(m)}\epsilon_{\alpha^\prime\beta^\prime}^{(m)}
     &=&{1\over 2}[{\cal P}_{\alpha\alpha^\prime}{\cal P}_{\beta\beta^\prime}+{\cal P}_{\alpha\beta^\prime}{\cal P}_{\beta\alpha^\prime}]
     -{1\over 3}{\cal P}_{\alpha\beta}{\cal P}_{\alpha^\prime\beta^\prime},\\
\nonumber
   \sum_m \epsilon_{\alpha\beta\rho}^{(m)}\epsilon_{\alpha^\prime\beta^\prime\rho^\prime}^{(m)}  
      &=&{1\over 6}({\cal P}_{\alpha\alpha^\prime}{\cal P}_{\beta\beta^\prime}{\cal P}_{\rho\rho^\prime}
              +{\cal P}_{\alpha\alpha^\prime}{\cal P}_{\beta\rho^\prime}{\cal P}_{\beta\rho^\prime}
              +{\cal P}_{\alpha\beta^\prime}{\cal P}_{\beta\alpha^\prime}{\cal P}_{\rho\rho^\prime}\\
\nonumber
        &~&~~~~+{\cal P}_{\alpha\beta^\prime}{\cal P}_{\beta\rho^\prime}{\cal P}_{\rho\alpha^\prime}                                                               
              +{\cal P}_{\alpha\rho^\prime}{\cal P}_{\beta\beta^\prime}{\cal P}_{\rho\alpha^\prime}
              +{\cal P}_{\alpha\rho^\prime}{\cal P}_{\beta\alpha^\prime}{\cal P}_{\rho\beta^\prime})\\
\nonumber
       &-&{1\over 15}({\cal P}_{\alpha\beta}{\cal P}_{\rho\alpha^\prime}{\cal P}_{\beta^\prime\rho^\prime}
              +{\cal P}_{\alpha\beta}{\cal P}_{\rho\beta^\prime}{\cal P}_{\alpha^\prime\rho^\prime}
              +{\cal P}_{\alpha\beta}{\cal P}_{\rho\rho^\prime}{\cal P}_{\alpha^\prime\beta^\prime}\\
\nonumber
        &~&~~~~+{\cal P}_{\alpha\rho}{\cal P}_{\beta\alpha^\prime}{\cal P}_{\beta^\prime\rho^\prime}
                 +{\cal P}_{\alpha\rho}{\cal P}_{\beta\beta^\prime}{\cal P}_{\alpha^\prime\rho^\prime}
                 +{\cal P}_{\alpha\rho}{\cal P}_{\beta\rho^\prime}{\cal P}_{\alpha^\prime\beta^\prime}\\               
        &~&~~~~+{\cal P}_{\beta\rho}{\cal P}_{\alpha\alpha^\prime}{\cal P}_{\beta^\prime\rho^\prime}
                 +{\cal P}_{\beta\rho}{\cal P}_{\alpha\beta^\prime}{\cal P}_{\alpha^\prime\rho^\prime}
                 +{\cal P}_{\beta\rho}{\cal P}_{\alpha\rho^\prime}{\cal P}_{\alpha^\prime\beta^\prime}).
\end{eqnarray}
Using these relations, the matrix elements $<0|(c\bar c)_{V-A}|{}^3D_J>$
can be calculated out,
\ba
<0|(c\bar c)_{V-A}|{}^3D_1>&=&\epsilon_\mu\frac{20\sqrt{3}}{\sqrt{2\pi}}
        \frac{R_D^{\prime\prime}(0)}{M^3\sqrt{M}},\\
<0|(c\bar c)_{V-A}|{}^3D_2>&=&0,\\
<0|(c\bar c)_{V-A}|{}^3D_3>&=&0.
\ea
Adopting this results, we obtain the partial width for color-singlet
$\delta^c_1$ production in $b$ decays,
\ba
\Gamma(b\to c\bar c({}^3D_1^{(1)})+s,d\to\delta^c_1+X)&=&
        \frac{25}{6\pi^2}\frac{|R_D^{\prime\prime}(0)|^2}{M^5}M_b^3(2C_+-C_-)^2
        (1+\frac{2M^2}{M_b^2})(1-\frac{M^2}{M_b^2})^2\\
        &=&\frac{5}{9}\frac{\< {\cal O}_8^{\delta^c_1}({}^3D_1)\> }{M_c^6}
        (2C_+-C_-)^2(1+\frac{8M_c^2}{M_b^2})\hat\Gamma_0.
\ea

The color-octet contributions to $\delta^c_J$ production in $b$ decays are
carried out by using the same way\cite{b-swave}
\ba
\Gamma(b\to (c\bar c)_8+s,d\to\delta^c_1+X)&=&\big (
        \frac{\< {\cal O}_8^{\delta^c_1}({}^3S_1)\> }{2M_c^2}+
        \frac{\< {\cal O}_8^{\delta^c_1}({}^3P_1)\> }{M_c^4}\big )
        (C_++C_-)^2(1+\frac{8M_c^2}{M_b^2})\hat\Gamma^0,\\
\Gamma(b\to (c\bar c)_8+s,d\to\delta^c_2+X)&=&\big (
        \frac{\< {\cal O}_8^{\delta^c_2}({}^3S_1)\> }{2M_c^2}+
        \frac{\< {\cal O}_8^{\delta^c_2}({}^3P_1)\> }{M_c^4}\big )
        (C_++C_-)^2(1+\frac{8M_c^2}{M_b^2})\hat\Gamma^0,\\
\Gamma(b\to (c\bar c)_8+s,d\to\delta^c_3+X)&=&\big (
        \frac{\< {\cal O}_8^{\delta^c_3}({}^3S_1)\> }{2M_c^2}+
        \frac{\< {\cal O}_8^{\delta^c_3}({}^3P_1)\> }{M_c^4}\big )
        (C_++C_-)^2(1+\frac{8M_c^2}{M_b^2})\hat\Gamma^0.
\ea
Adding the color-singlet and color-octet contributions together, for $\delta^c_1$
production, the patial width is
\ba
\nonumber
\Gamma(b\to \to\delta^c_1+X)&=&(1+\frac{8M_c^2}{M_b^2})\hat\Gamma^0
        \big [\frac{5\< {\cal O}_1^{\delta^c_1}({}^3D_1)\> }{9M_c^6}(2C_+-C_-)^2\\
        &~&+(
        \frac{\< {\cal O}_8^{\delta^c_1}({}^3S_1)\> }{2M_c^2}+
        \frac{\< {\cal O}_8^{\delta^c_1}({}^3P_1)\> }{M_c^4} )(C_++C_-)^2\big ].
\ea

\section{Numerical estimation and discussion of the detection}

In the numerical calculations, we take the input parameters as
\beq
M_b\approx M_B=5.3GeV,~~~M=2M_c=3.8GeV,~~~\alpha_s(M_b)=0.20,~~~\alpha_s(M_z)=0.116,
\eeq
and then,
\beq
C_+(M_b)=0.87,~~~C_-(M_b)=1.34.
\eeq
By using the standard method \cite{b-pwave}, we obtain the inclusive branching
fractions for $\delta^c$ and $\delta^c_J$ 
\ba
BR(B\to \delta^c+X)&=&0.76\frac{\< {\cal O}_8^{\delta^c}({}^1S_0)\> }{M_c^3},\\
BR(B\to \delta^c_1+X)&=&0.019\frac{\< {\cal O}_1^{\delta^c_1}({}^3D_1)\> }{M_c^7}
        +1.03\big (\frac{\< {\cal O}_8^{\delta^c_1}({}^3S_1)\> }{2M_c^3}+
        \frac{\< {\cal O}_8^{\delta^c_1}({}^3P_1)\> }{M_c^5} \big ),\\
BR(B\to \delta^c_2+X)&=&
        1.7\big (\frac{\< {\cal O}_8^{\delta^c_1}({}^3S_1)\> }{2M_c^3}+
        \frac{\< {\cal O}_8^{\delta^c_1}({}^3P_1)\> }{M_c^5} \big ),\\
BR(B\to \delta^c_3+X)&=&
        2.4\big (\frac{\< {\cal O}_8^{\delta^c_1}({}^3S_1)\> }{2M_c^3}+
        \frac{\< {\cal O}_8^{\delta^c_1}({}^3P_1)\> }{M_c^5} \big ),
\ea
where the heavy quark spin symmetry relations for the color-octet matrix 
elements have been used,
\ba
\< {\cal O}_8^{\delta^c_J}({}^3S_1)\> &\approx &\frac{2J+1}{3}
        \< {\cal O}_8^{\delta^c_1}({}^3S_1)\> ,\\
\< {\cal O}_8^{\delta^c_J}({}^3P_1)\> &\approx &\frac{2J+1}{3}
        \< {\cal O}_8^{\delta^c_1}({}^3P_1)\> .
\ea
>From the above results, we can see that the color-octet contributions are
crucial important to $D$-wave charmonium production in $B$ decays.
The color-singlet contribution is not vanishing only in the case of $J=1$.
Furthermore, in that case, the color-singlet contribution is too small
compared with the color-octet contributions, because the color-octet matrix
elements are at the same order in $v^2$ as the color-singlet matrix element 
but the short-distance coefficients in the color-octet terms are about $50$
times larger than that in the  color singlet term. 
So, the color-octet contributions are much more important than the color 
singlet contribution.

The values of color-singlet matrix elements 
are gotten from the potential model calculation 
$|R_D^{\prime\prime}(0)|^2=0.015 GeV^7$\cite{potential}.
Using the relations Eqs.(8-9), we get
\beq
\< {\cal O}_1^{\delta^c}({}^1D_2)\> =0.27 GeV^7,~~~
        \< {\cal O}_1^{\delta^c_1}({}^3D_1)\> =0.16 GeV^7.
\eeq
We estimate the size of the color-octet matrix elements by using the naive
NRQCD velocity scaling rules,
\ba
\frac{\< {\cal O}_8^{\delta^c_1}({}^3S_1)\> }{M_c^3}&\approx &
        \frac{\< {\cal O}_8^{\delta^c_1}({}^3P_1)\> }{M_c^5}\approx
        \frac{\< {\cal O}_1^{\delta^c_1}({}^3D_1)\> }
        {M_c^7}=0.0018,\\
\frac{\< {\cal O}_8^{\delta^c}({}^1S_0)\> }{M_c^3}&\approx &
        \frac{\< {\cal O}_1^{\delta^c}({}^1D_2)\> }
        {M_c^7}=0.0030.
\ea
Here, these equation only make sense in the estimate of order of magnitudes 
(we may also use another estimate for these relations as in Ref.\cite{gluon},
in that case, the results of this paper will be enhanced about one order).
Adopting these matrix elements values, the predicted branching fractions for
$D$-wave charmonium states are,
\ba
BR(B\to \delta^c+X)&=&0.23\%,\\
BR(B\to \delta^c_1+X)&=&0.28\%,\\
\label{d2}
BR(B\to \delta^c_2+X)&=&0.46\%,\\
BR(B\to \delta^c_3+X)&=&0.65\%.
\ea
We can easily find that the relative production rates predicted above are 
$\delta^c:\delta^c_1:\delta^c_2:\delta^c_3=2.5:3:5:7$. If we do
not take into account the color-octet mechanism the relative rates would be
$\delta^c:\delta^c_1:\delta^c_2:\delta^c_3=0:1:0:0$.

Among the three triplet states of $D$-wave charmonium, $\delta^c_2$ is the
most prominant candidate to be discovered firstly. It is a narrow resonance,
and its branching fraction of the decay mode $\jpsi \pi^+\pi^-$ is estimated 
to be \cite{z0}
\beq
B(\delta^c_2\to \jpsi \pi^+\pi^-)\approx 0.12,
\eeq
which is only smaller than that of $B(\psi^{\prime}\to \jpsi \pi^+\pi^-)$ by only 
a
factor of $3$. Therefore the decay mode $\delta^c_2\to \jpsi \pi^+\pi^-$ could 
be
observable if the production rate of $\delta^c_2$ is of the same order as
$\psi^\prime$.
Comparing the predicted production rate of $2^{--}$ $D$-wave charmonium (\ref{d2})
with that of $\psi^\prime$\cite{b-swave}, we can
see that the former has the same amount of that of the latter.
At {\bf CESR}, the {\bf CLEO} collobration have observed strong signals of
$\psi^\prime$ by $\jpsi\pi\pi$ triggering\cite{cleo}.
Recently, at {\bf LEP}, the {\bf OPAL} collaboration have also detected $\psi^\prime$
by the same trigger with higher statistics\cite{opal}.
Therefore, we would expect the detection of $2^{--}$ $D$-wave charmonium state at these two
machines will soon be obtained.

The other two states of the spin-triplet, $\delta^c_1$ and $\delta^c_3$ are
above the open channel threshold and are not narrow, and therefore are difficult
to detect. However, the analysis of $\jpsi\pi^+\pi^-$ spectrum in $Z^0$ decays
performed by the {\bf OPAL} collaboration shows some events above the background
around $3.77GeV$ besides the events peak at $\psi^\prime(3686)$\cite{opal}. If these events
are finally confirmed to be associated with the $1^{--}$ $D$-wave charmonium state
$\psi(3770)$, they may be mostly from $b$ decays. Moreover, $\psi(3770)$ could
also be seen in a $D\bar{D}$ (charmed meson pair) final state.

At the {\bf Tevatron}, about $10^9$ $b$ quarks are produced with an accumulated
luminosity of $100Pb^{-1}$, which implies more than $10^6$ $D$-wave charmonium
particles could be produced. The estimated combined branching ratio $B(\delta^c_2\to \jpsi\pi^+\pi^-,
\jpsi\to \mu^+\mu^-)\approx 7\times 10^{-3}$ shows that
more than $10^3$ events of $2^{--}$ $D$-wave charmonium will be detected at
this collider. Furthermore, the combined branching ratio
$B(\delta^c\to {}^1P_1\gamma,{}^1P_1\to\jpsi \pi^0,\jpsi\to \mu^+\mu^-)$
has been estimated to be about $10^{-4}$\cite{2d1}. This shows that more than $10^2$
events of this particle can also be detected.

\section{Summay}

We have calculated $D$-wave charmonium production rates 
in $B$ decays in this paper.
We find that the inclusion of color-octet production mechanism can help us to
obtain the $D$-wave states production rates at an observable level. 
In our calculations, the values of the color-octet matrix elements are taken 
by assuming that the NRQCD velocity scaling rules are valid in these cases. 
Practically, these matrix elements can be determined by fitting the theretical
prediction rates to the experimental data.
In \cite{z0}\cite{gluon}, these matrix elements also appear in the calculations.
If $D$-wave charmonium states are detected in the future, all these matrix
elements can be extracted from experimental data, and the compararison of
the results from different experiments will provide another important
test of the universality of the color-octet matrix elements, the NRQCD 
velocity scaling rules, and further the NRQCD factorization formalism.


\end{document}